# LLMapReduce: Multi-Level Map-Reduce for High Performance Data Analysis

Chansup Byun, Jeremy Kepner, William Arcand, David Bestor, Bill Bergeron, Vijay Gadepally, Matthew Hubbell, Peter Michaleas, Julie Mullen, Andrew Prout, Antonio Rosa, Charles Yee, Albert Reuther

MIT Lincoln Laboratory, Lexington, MA, U.S.A *Abstract*— The map-reduce parallel programming model has become extremely popular in the big data community. Many big data workloads can benefit from the enhanced performance offered by supercomputers. LLMapReduce provides the familiar map-reduce parallel programming model to big data users running on a supercomputer. LLMapReduce dramatically simplifies map-reduce programming by providing simple parallel programming capability in one line of code. LLMapReduce supports all programming languages and many schedulers. LLMapReduce can work with any application without the need to modify the application. Furthermore, LLMapReduce can overcome scaling limits in the map-reduce parallel programming model via options that allow the user to switch to the more efficient single-program-multiple-data (SPMD) parallel programming model. These features allow users to reduce the computational overhead by more than 10x compared to standard map-reduce for certain applications. LLMapReduce is widely used by hundreds of users at MIT. Currently LLMapReduce works with several schedulers such as SLURM, Grid Engine and LSF.

## I. INTRODUCTION

Large scale computing is currently dominated by four ecosystems: supercomputing, database, enterprise, and big data [1]. Each of these ecosystems has its strengths.

The supercomputing ecosystem provides the highest performing computing capabilities in the world via a range of well-established, highly-optimized, specially-designed high performance technologies. Among these technologies are high performance messaging libraries (e.g., MPI [2, 3]) that utilize high performance interconnects (e.g., InfiniBand [4], IBM Blue Gene interconnects [5], Cray interconnects [6]), High performance math libraries (e.g., BLAS [7, 8], LAPACK [9], ScaLAPACK [10]) designed to exploit special processing hardware, high performance parallel files systems (e.g., Lustre [11], GPFS [12]) and high performance schedulers (e.g., LSF [13], GridEngine [14, 15, 16], SLURM [17, 18]). Combined, these technologies consistently achieve near peak speedups on well-established benchmarks (e.g., HPC Challenge [19]) and many applications. On the largest systems in the world these speedups can be in the millions.

Relational or SQL (Structured Query Language) database ecosystems [20, 21] have been the de facto interface to databases since the 1980s and are the backbone of electronic transactions around the world. Enterprise ecosystems often exploit virtualization technologies (e.g., VMware) to deliver a vast range of web services such as e-mail, calendar, document sharing, videos, and product information. Big data ecosystems represent non-traditional, relaxed consistency, triple store databases which provide high performance on commodity computing hardware to I/O intensive data mining applications with low data modification requirements. These databases, which are the backbone of many web companies, include Google Big Table [22], Amazon Dynamo [23], Cassandra [24], and HBase [25].

The MIT Lincoln Laboratory Supercomputing Center (LLSC) provides supercomputing capabilities to over 1000 users at MIT [26, 27]. Increasingly, these users require capabilities that are found in all four ecosystems. LLSC has developed the MIT SuperCloud environment that allows all four ecosystems to run on the same hardware without sacrificing performance [1]. The MIT SuperCloud has spurred the development of a number of cross-ecosystem innovations in high performance databases [28, 29], database management [30], database federation [31, 32, and 33], data analytics [34], data protection [35], and system monitoring [36, 37].

One of the most impactful MIT SuperCloud innovations has been the LLMapReduce environment that is used by a large fraction of the LLSC user base to perform map-reduce computations on LLSC supercomputers. The map-reduce parallel programming model is one of the oldest parallel programming models. The map-reduce name derives from the "map" and "reduce" functions found in Common Lisp since the 1990s. Popularized by Google [38] and Apache Hadoop [39], map-reduce has become a staple technology of the ever-growing big data community. Traditionally, the big data community has relied on inexpensive clusters for the majority of its computational needs. Recently, it has become apparent that many big data workloads can benefit from the superior performance and ease-of-use offered by supercomputers [29, 30]. Many LLSC users are recent graduates and have prior experience with the map-reduce programming model. It became vital to LLSC to develop an environment that would allow these users to leverage their prior map-reduce knowledge in a way that would allow them take advantage of the superior performance offered by a supercomputing ecosystem.

Map-reduce popularity stems from its simplicity. A map-reduce program consists of two parts: a map that runs the same program on many inputs (usually files) and produces one output for each input and a reduce that runs another program on the map outputs to produce a single final result. This style of programming can be easily implemented in any mature supercomputing scheduler via a simple submission script. In addition, supercomputing schedulers have the advantage of being programming language agnostic.

The Apache Hadoop map-reduce implementation provides a number of benefits such as automatic parallelization and fault-tolerant features for Java programmers [40]. Although the support has been extended to other languages such as Python [41], it still requires significant modification of programmer code to use.

Scientists and engineers at MIT LLSC must work with codes that are written in many languages. The LLSC team developed and deployed LLMapReduce on its supercomputers systems [28], using the Lustre central storage system instead of the Apache Hadoop distributed filesystem (HDFS) [39]. LLMapReduce can launch any program in any language on any supercomputers with a standard scheduler. Workloads can be distributed in a block or cyclic fashion to improve load balancing. Since the initial deployment in 2012, LLMapReduce has added a variety of features to reduce the performance overheads associated with map-reduce programming. One of the most compelling features is the LLMapReduce multiple-input-multiple-output (MIMO) mode that transparently morphs a user's program from a map-reduce parallel programming model to the highly efficient single-program-multiple data (SPMD) data model that is widely used in the supercomputing ecosystem. Thus, a user gets the familiarity of map-reduce and the performance of SPMD. Using this feature often results in 10x increase in performance over standard map-reduce. Currently, LLMapReduce works with most schedulers including SLURM [17, 18], the open and commercial distributions of Grid Engine [14, 15, 16] and IBM Platform LSF [13].

The rest of the paper is organized as follows. Section II presents an overview of LLMapReduce. Section III describes specific use case examples of LLMapReduce. Section IV presents various performance results. The work is summarized in section V.

## II. LLMAPREDUCE

The map-reduce parallel programming model is one of the simplest of all parallel programming models; for many programmers it is easier to learn than message passing or distributed arrays. The map-reduce parallel programming model consists of two user written programs: a Mapper and a Reducer. The input to the Mapper is a file, and the output is another file. The input to the Reducer is the set of Mapper output files. The output of Reducer is a single file. Launching consists of starting many Mapper programs each with a different input file. When the Mapper programs all have completed, the Reduce program is run on the Mapper outputs.

LLMapReduce has been deployed on supercomputer systems by utilizing a centralized high-performance parallel filesystem such as the Lustre filesystem [11]. Because most supercomputers run a scheduler, LLMapReduce is designed to use the existing schedulers to manage its workloads. LLMapReduce was originally written to work with the open source Grid Engine [15, 16], and more recently it has been extended to work with SLURM [17, 18] and LSF [13]. This allows that LLMapReduce presents a single scheduler-neutral API interface to hide the incompatibility among the schedulers.

LLMapReduce assumes that users will have their data already partitioned into data files. Such segmentation is natural for many application areas. For example, when collecting data from various sensors, they are often collected in a large number of segmented files instead of one large, holistic file. This allows users to deploy their applications rapidly and efficiently. LLMapReduce identifies the input files to be processed by scanning a given input directory or reading a list from a given input file as shown in the step 1 in Fig. 1. LLMapReduce generates all the necessary temporary files under the directory, .MAPRED.PID, where the PID is the process identification number.

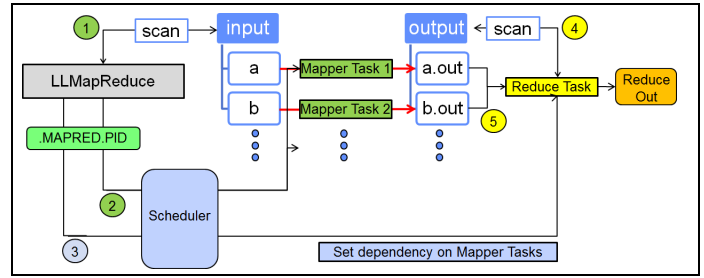

Fig. 1. A schematic diagram showing how LLMapReduce works. The Map Reduce process identifies input files which are used to generate an array job with the help of the HPC scheduler. By setting a dependency between the mapper and reducer jobs, the output of completed jobs are passed through a reducer to generate the final result.

Many filesystems operate best when the number of files per directory is less than 10,000. LLMapReduce users can build a nested call to LLMapReduce for processing whole hierarchies of data. Then, by accessing the scheduler at step 2, it creates an array of many tasks, called an array job, which is denoted as "Mapper Task 1", "Mapper Task 2", and so on.

Once the array job is created and dispatched for execution, each input file will be processed by one of the tasks with the specified application at the command line, noted as "Mapper" in Fig. 1. The application can be any type of executable written in any language, such as a shell script, a Java program or a MATLAB script. In addition, there is an option to do further processing on the results, if there are any, by creating a dependent task at step 3. This is denoted as the "Reduce Task" in Fig. 1. The reduce task will wait until all the mapper tasks are completed by setting a job dependency between the mapper tasks and reduce task. The reduce application is responsible to scan the output files from the mapper tasks at step 4 and to process them into the final results at step 5.

The map application of LLMapReduce requires two input arguments: one for the input filename and the other for the output filename. Subsequently, the reducer application takes two arguments as input, which are the directory path where the results of the map tasks reside and the output filename for the reduce result. The reduce application may use the input path to scan and read the output generated by the map tasks.

The available options of LLMapReduce are shown in Fig. 2. The --np option can not only limit the total number of array tasks a job generates, but also calculates the number of data files per each array task to be assigned. It is possible for users to launch their applications with a number of data files that exceeded the limit that the scheduler job array could accommodate. The --ndata option allows users to define how many data files are to be assigned per each array task, which will override the --np option. The --ext option enables changing the default extension, "out", with the user-defined extension name. Along with the --ext option, the --delimiter option enables the change of the default, "." extension with a user-defined delimiter when adding the extension. The --distribution option enables changing how the input data is distributed among the given number of task processes; the default is the block distribution. The --exclusive option enables the use of entire compute nodes for jobs. By default, LLMapReduce will delete the .MAPRED.PID directory after the job is completed. However, users can keep the temporary directory for debugging purpose with the --keep=true option. LLMapReduce can also include some additional scheduler options when generating the job submission scripts with the --options option. This is handy when some data processing requires more memory than the standard allowance, for instance.

```
LLMapReduce --np=number_of_tasks        \
            --input=input_dir           \
            --output=output_dir         \
            --mapper=myMapper           \
            --reducer=myReducer         \
            --redout=output_filename    \
            --ndata=NdataPerTask        \
          --distribution=block|cyclic   \
            --subdir=true|false         \
            --ext=myExt                 \
          --delimiter=myExtDelimiter    \
            --exclusive=true|false      \
            --keep=true|false           \
            --apptype=mimo|siso         \
   --options=<scheduler_options_to_add>
```

Fig. 2. Available options of LLMapReduce.

### A. Data in a Hierachical Directory Structure

The --subdir option enables the processing of data files stored in a hierarchical directory structure as shown in Fig. 3. Hierarchical data storage is desirable for simplifying the organization of data and for enhancing the performance of the underlying filesystem. For example, a full listing of 100,000 files in a single directory on a Lustre filesystem can take longer than desired. For performance reason, it is recommended to distribute 100,000 files into multiple sub-directories. However, this will make the data processing inconvenient to users because users need to take care of the directory hierarchy when submitting their jobs for processing.

By defining the --subdir=true option with the top directory path of to the input data, as shown in Fig. 3, LLMapReduce will scan the input directory recursively and list all the files under the input directory as input data to the map process. In addition, LLMapReduce will duplicate the input data structure to the output directory. This option allows users to deal with a large number of input files intuitively and efficiently.

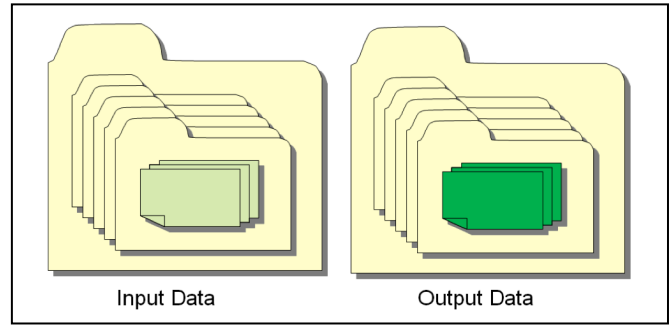

Fig. 3. The --subdir=true option replicates the directory hierachy structure between the input and output directory.

### B. Multi-Level Map Reduce

By default, LLMapReduce expects that the map application takes a single input and single output path (siso) as shown in Fig. 4(a). However, this will incur overhead associated with repeated startups of the map application. Some applications such as MATLAB codes can save significant overhead cost with the minor change of having the map application start only once and read many lines of input/output path pairs to process the given data as shown in Fig. 4(b). For this purpose the --apptype=mimo option will generate the input files for the modified map application that will read the input file with the multiple lines of input/output filename pairs. With the --apptype=mimo option, other applications with relatively small start-up time still benefit from reducing the latency overhead associated with the scheduler job launch mechanism.

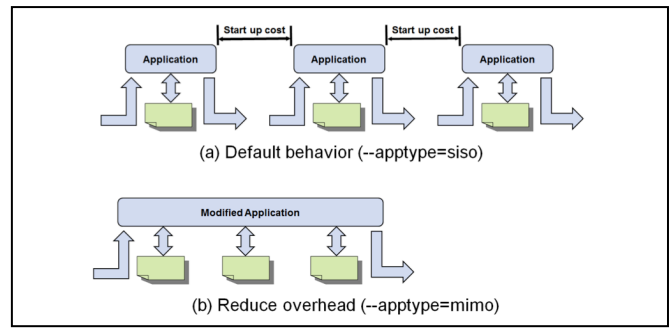

Fig. 4. Reducing overhead costs associated with repeated startups of the map application with the –apptype=mimo option.

This option is particularly useful for processing a large number of data files with relatively short running applications as compared to their startup time. In this case, the time saving from the option becomes significant. However, if map applications run for a long time compared to their startup time, the cost saving is less.

### III. USE CASES

In this section, two use cases are presented to demonstrate how LLMapReduce is used in practice. First, a MATLAB application that converts an RGB image into a gray-scale image is used with a small number of image files. It demonstrates the use of the Mapper and assigning multiple input files to one application execution. Next, a Java application that counts the number of unique words in the given text files illustrates the use of the Mapper and Reducer.

*A. MATLAB Application*

An image conversion function, called imageConvert(), is shown in Fig. 5. The function takes two arguments, the input and output image names. It reads in an RGB image file and converts it into a gray scale image. Then, it saves the gray scale image into the file of the output name. This satisfies the LLMapReduce API requirements.

To run the MATLAB program requires a wrapper script to receive the input and output file names that are provided to LLMapReduce. The wrapper script will execute MATLAB with the corresponding input and output files when dispatched for execution by the scheduler. An example wrapper script is shown in Fig. 6.

```
function imageConvert(inFile,outFile)
I=imread(inFile); J=rgb2gray(I);
dicomwrite(J,outFile);
```

Fig. 5. A MATLAB function, imageConvert(), to convert an RGB image into a gray scale image.

```
#!/bin/bash
cat<<EOF|matlab -nodisplay -singleCompThread
inFile='$1'; outFile='$2';
imageConvert(inFile, outFile);
EOF
```

Fig. 6. An example wrapper script, MatlabCmd.sh, for the imageConvert() function.

In the above wrapper script, the variables, $1 and $2, are the two input arguments (input and output file names), which are provided by LLMapReduce. The script will execute the imageConvert() function with MATLAB when it is called by the run script, which is generated by LLMapReduce. With the wrapper script, the image conversion job can be launched with one line of the LLMapReduce command as shown in Fig. 7. In this case, each input image file in the input directory becomes an array task of an array job automatically generated by LLMapReduce. The resulting gray images are saved in the output directory as specified.

```
$ LLMapReduce --mapper MatlabCmd.sh \
              --input input --output output
```

Fig. 7. An example Map Reduce job with LLMapReduce.

When the LLMapReduce command is called, temporary files are created in the .MAPRED.PID directory, where PID is the process identification (PID) number of the LLMapReduce command. These temporary files are generated for the specific scheduler being used on a particular supercomputer. The files for one job submission script and a number of run scripts for all array tasks (one per array task) is shown in Figs. 8 and 9, respectively. The job submission script shown in Fig. 8 is written for the open source Grid Engine scheduler, which has a number of options specific to the scheduler. LLMapReduce hides the scheduler-specific job submission script from users and, therefore, provides a scheduler-neutral API. The -t 1-M option specifies an array job of M tasks, starting from 1 to M with an increment of one. M is determined by LLMapReduce, which is the number of input image files in the input directory. Each array task keeps its own log file, uniquely named with its job and task numbers. If there is any standard output, it goes into these log files.

```
#!/bin/bash
#$ -terse -cwd -V -j y -N MatlabCmd.sh
#$ -l excl=false -t 1-M
#$ -o .MAPRED.1120/llmap.log-$JOB_ID-$TASK_ID
./.MAPRED.1120/run_llmap_$SGE_TASK_ID
```

Fig. 8. An example job submission script generated for the Grid Engine scheduler.

```
$ cat .MAPRED.1120/run_llmap_1 (for task 1)
#!/bin/bash
export PATH=${PATH}:.
MatlabCmd.sh input/image_1.jpg \
             output/image_1.jpg.out
. . .
$ cat .MAPRED.1120/run_llmap_M (for task M)
#!/bin/bash
export PATH=${PATH}:.
MatlabCmd.sh input/image_M.jpg \
             output/image_M.jpg.out
```

Fig. 9. A number of run scripts for all array tasks generated by the LLMapReduce.

The run script for each array task is generated to feed one input and one output argument to the wrapper script shown in Fig. 6. This meets the LLMapReduce API requirement. As mentioned above, LLMapReduce generates M run scripts, one run script per each array task. As shown in Fig. 9, the output file name is determined by the name of the input file with the default extension, ".out".

The example shown in Fig. 7 has an issue if there are a large number of input files in the given input directory. This can exceed the scheduler limit for how many tasks a job array can have. For example, the default maximum number of array tasks for an array job is 75,000 for the open source Grid Engine scheduler. To handle this case, the --np option can be used to specify how many array tasks LLMapReduce should create. For example, if the --np=100 option is used, only 100 array tasks are created and each array task will process a block of the total input data instead of a single data file. The block size is determined by LLMapReduce.

```
$ LLMapReduce --mapper MatlabCmdMulti.sh \
         --input input --output output \
         --np N --apptype mimo --ext gray
```

Fig. 10. An example map-reduce job to eliminate the overhead cost associated with the multiple execution of an application in the default processing model.

With the --np option, a large number of input data files can be handled easily. This approach executes the application multiple times (as many as the number of input data files). There is a significant overhead cost associated with the repeated startups of the application, especially in complex programming environments such as MATLAB. One way to greatly reduce the overhead cost is to launch the application once and process all the data assigned to each array task. This requires modifying the wrapper script shown in Fig. 6 in addition to modifying the job submission script and the run scripts shown in Figs. 8 and 9. This feature can be invoked by the command shown in Fig. 10. The command specifies a wrapper script, MatlabCmdMulti.sh, to handle the multiple lines of input and output file lists, created with the --apptype=mimo option. Also, with the --ext=gray option, the extension is renamed as ".gray" instead of the default, ".out". An example wrapper script, MatlabCmdMulti.sh, for the --apptype=mimo option is presented in Fig. 11. This script reads in the input and output file names from the generated file provided by the run scripts, which are also generated by LLMapReduce. The Fig. 11 script launches the application (MATLAB in this case) once and processes all the data, assigned by a dynamically generated file.

```
cat<<EOF|matlab -nodisplay -singleCompThread
inFile='$1'; fid=fopen(inFile);
tline=fgets(fid);
while ischar(tline)
  myStr=strsplit(tline);
  indata=deblank(myStr{1});
  outdata=deblank(myStr{2});
  imageConvert(indata, outdata);
  tline=fgets(fid);
end
fclose(fid);
EOF
```

Fig. 11. A wrapper script, MatlabCmdMulti.sh, to accept multiple lines of input and output file names with the --apptype=mimo option.

```
$ cat .MAPRED.2188/run_llmap_1 (for task 1)
#!/bin/bash
export PATH=${PATH}:.
MatlabCmdMulti.sh ./.MAPRED.2188/input_1
. . .
$ cat .MAPRED.2188/run_llmap_N (for task N)
#!/bin/bash
export PATH=${PATH}:.
MatlabCmdMulti.sh ./.MAPRED.2188/input_N
```

Fig. 12. A number of run scripts for all array tasks generated by LLMapReduce.

With the --apptype=mimo option, LLMapReduce generates a job submission script similar to Fig. 8. However, different run scripts are generated, named as run_llmap_x, which is shown in Fig. 12. The number x ranges from 1 to N, where N is the number of array tasks defined by the --np option. In these run scripts, the wrapper script takes one input file, which is automatically generated by LLMapReduce. The input files named as input_x, which are also generated by LLMapReduce, list the input and output file names.

### B. A Java Application

A common map-reduce example is a word frequency counting application. We've chosen an example by Swartz [42] instead of the Hadoop Word Count example for simplicity. In the example, the WordFreqCmd.java code requires three command line inputs: input, output, and a reference file. The reference file contains a list of words to be ignored for word counting. To comply with the LLMapReduce API requirement, a wrapper script, WordFreqCmd.sh, has been created as shown in Fig. 13. In the wrapper script, the variables, $1 and $2, represent the input and output files, respectively.

```
#!/bin/bash
java WordFrequencyCmd $1 $2 textignore.txt
```

Fig. 13. A wrapper script for the WordFreqCmd.java code.

Another wrapper script, ReduceWordFreqCmd.sh, is used to execute the ReduceWordFrequencyCmd.java code to collect the map process results as shown in Fig. 14. The reduce code scans the map results in the output directory and merges the results into a single file. The reducer wrapper script expects two inputs: the first argument ($1) is the location of the map process results and the second argument ($2) is the output name of the reduce application. Both inputs are provided by LLMapReduce.

```
#!/bin/bash
java ReduceWordFrequencyCmd $1 $2
```

Fig. 14. A wrapper script for the ReduceWordFreqCmd.java code.

With these two wrapper scripts, a map-reduce job for the word frequency count can be launched by using LLMapReduce as shown in Fig. 15. In this example, to distribute the input data among the given number of array tasks in a cyclic fashion, the --distribution=cyclic option is used. As mentioned in the previous MATLAB example, the map-reduce job launched by the command in Fig. 15 also incurs the computational overhead associated with the multiple startups of the word count application. The run script for the reduce task is submitted as a dependent job to the mapper job that only uses one task. The java application, ReduceWordFrequencyCmd, scans the results in the given directory (output) and merges them into a single file (default name: llmapreduce.out).

```
$ LLMapReduce --np 3
            --mapper WordFreqCmd.sh \
            --reducer ReduceWordFreqCmd.sh \
            --input input --output output \
            --distribution cyclic
```

Fig. 15. A Map Reduce job for word frequency count using LLMapReduce.

As with the MATLAB example, the overhead can be reduced by launching with the --apptype=mimo option as shown in Fig. 16. This option also requires application modification in order to read in multiple lines of input and output pairs provided by LLMapReduce. The modified WordFreqCmdMulti.sh script for the --apptype =mimo option is shown in Fig. 17. It is similar to the script shown in Fig. 13 but executes a modified Java application called WordFrequencyCmdMulti.

```
$ LLMapReduce --np 3
   --mapper WordFreqCmdMulti.sh \
   --reducer ReduceWordFreqCmd.sh \
   --input input --output output \
   --apptype mimo
```

Fig. 16. A Map Reduce job for word frequency count using LLMapReduce with overhead cost reduction.

```
#!/bin/bash
java WordFrequencyCmdMulti $1 $2 \
                         textignore.txt
```

Fig. 17. A modified wrapper script for the WordFreqCmdMulti.java code.

The modified Java code, WordFrequencyCmdMulti.java, has some additional lines, which reads in multiple lines of the input and output filename pairs. The input to the modified Java code is automatically generated by LLMapReduce. The original section of the code processes the given input file and writes the results to the given output file. As a result, the Java code is invoked only once and processes all the input data assigned to its array task.

## IV. PERFORMANCE

In this section, the performance benefits of the --apptype=mimo (MIMO) option are presented. Two use cases from the previous section in addition to a user MATLAB application are used. Furthermore, the behavior of the three different options (DEFAULT, BLOCK, and MIMO) are compared by varying the number of processes and the number of input data files.

The toy examples described previously are small in terms of number of input data files. The MATLAB application converts 6 images over 2 array tasks. The Java application counts word frequency of 21 text files over 3 array tasks. The map-reduce jobs were executed with the BLOCK and MIMO options, and the total processing time was measured. The speed up is calculated by the ratio between the time with the BLOCK option and the time with the MIMO option. The results are presented in Table 1. Although there are only a small number of data files assigned to each array task, both examples show speed up with the MIMO option.

TABLE I. SPEED UP WITH TOY EXAMPLES

| Example | Type | Speed up |
|---|---|---|
| Matlab | Multiple app launches (BLOCK) | 1 |
| | Single app launch (MIMO) | 2.41 |
| Java | Multiple app launches (BLOCK) | 1 |
| | Single app launch (MIMO) | 2.85 |

A performance study with a real user MATLAB application has been performed and the results are presented in Table 2. The MATLAB application does image processing, and the image files were distributed to 256 array tasks. The number of input data files was 43,580 in this example. As MATLAB takes relatively significant time to launch as compared to other programs, the performance difference was significant. By using the MIMO option, the map-reduce job was able to run almost 12 times faster than the BLOCK option.

TABLE II. SPEED UP WITH A REAL WORLD APPLICATION

| Example | Type | Speed up |
|---|---|---|
| Matlab | Multiple app launches (BLOCK) | 1 |
| | Single app launch (MIMO) | 11.57 |

For the scalability study, three different LLMapReduce options (DEFAULT, BLOCK, and MIMO) were used with a MATLAB code that reads in a list of square matrices and multiplies the matrices. 512 input data files were created and run with various numbers of concurrent array tasks (processes), ranging from 1, 2, 4, 8, 16, 32, 64, 128, and 256 for the three different options. The results are presented in Figs. 18 and 19.

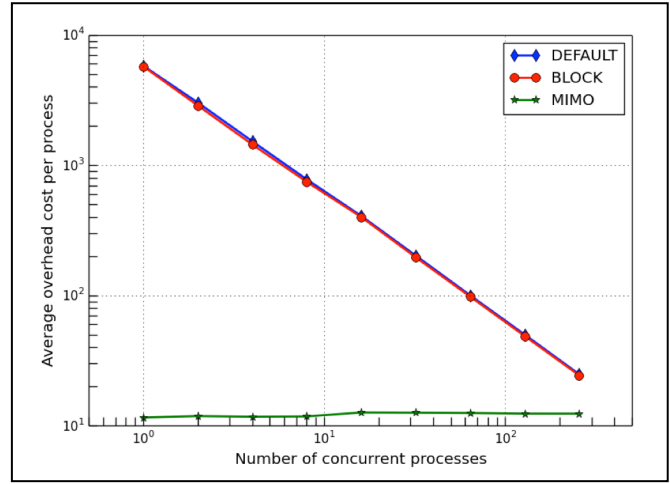

Fig. 18. The computational overhead cost when varying the number of concurrent array tasks (processes), which changes the number of the input data files per array task.

Fig. 18 shows the computational overhead associated with the cost of multiple application start-ups when dealing with more than one input data file per array task. While the cases for the DEFAULT and BLOCK options show that the average overhead cost per array task decreases linearly as the number of array tasks is increased, the overhead cost for the MIMO option remains relatively flat. As far as the overhead cost is concerned, both DEFAULT and BLOCK options show similar overhead, although the BLOCK option shows slightly smaller cost. The MIMO overhead cost is significantly smaller than those of the other two options. Thus the gap in the overhead cost between the MIMO and the other two options becomes significant when each array task processes a large number of data files. Fig. 18 clearly shows the benefits of the MIMO option when dealing with a large number of input data files per array task.

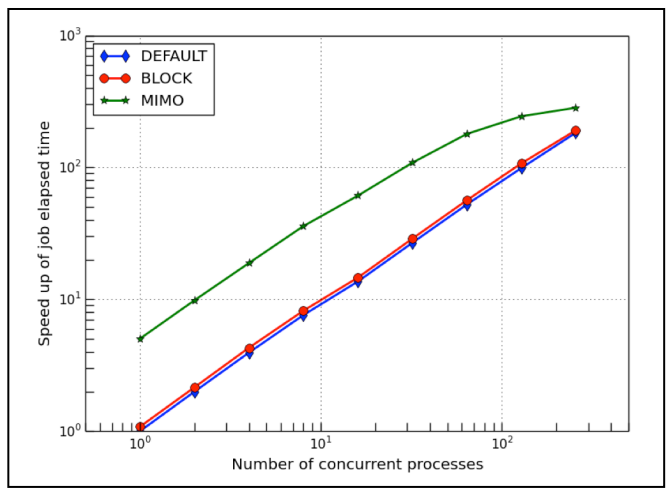

Fig. 19. The speed-up of job elapsed times with respect to the default job elapsed time with one compute process when varying the number of concurrent array tasks (processes), which in turn changes the number of the input data files per array task.

Fig. 19 shows the speed-up based on job elapsed times for the three different options with varying number of array tasks (processes). The speed-up is calculated by the ratio between the DEFAULT job elapsed time obtained with one array task and the other job elapsed times. Throughout all the numbers of the concurrent array tasks, the MIMO option performed the best, consistently outperforming the other two options. The BLOCK option performed slightly better than the DEFAULT option but the difference is marginal. As the number of concurrent array tasks is increased, the overhead cost per array task diminishes. If each array task processes only one data file, the results of all three options will converge at the same point.

## V. SUMMARY

LLMapReduce has been developed and deployed on the supercomputers to support scientists and engineers at MIT. With LLMapReduce, users can deploy their map-reduce style applications quickly on a supercomputer. LLMapReduce can work with any executable application without the need for any modifications. However, for improved performance, LLMapReduce provides an option to consolidate multiple input data files per array task as a single stream of input with minimal changes to the target application. This enables users to cut down the computational overhead associated with the cost of repeated application start-ups when dealing with more than one input data file per array task. With a small change in a sample MATLAB image processing application, we have observed approximately 12x speed up by reducing the overhead associated with the repeated application start-ups. Currently LLMapReduce works with handful of schedulers including SLURM, Grid Engine and LSF.